# New Approach for Prediction Pre-cancer via Detecting Mutated in Tumor Protein P53

Ayad Ghany Ismaeel

**Abstract**— Tumor protein P53 is believed to be involved in over half of human cancers cases, the prediction of malignancies plays essential roles not only in advance detection for cancer, but also in discovering effective prevention and treatment of cancer, till now there isn't approach be able in prediction the mutated in tumor protein P53 which is caused high ratio of human cancers like breast, Blood, skin, liver, lung, bladder etc. This research proposed a new approach for prediction pre-cancer via detection malignant mutations in tumor protein P53 using bioinformatics tools like FASTA, BLAST, CLUSTALW and TP53 databases worldwide. Implement and apply this new approach of prediction pre-cancer through mutations at tumor protein P53 shows an effective result when used more specific parameters/features to extract the prediction result that means when the user increase the number of filters of the results which obtained from the database gives more specific diagnosis and classify, addition that the detecting pre-cancer via prediction mutated tumor protein P53 will reduces a person's cancers in the future by avoiding exposure to toxins, radiation or monitoring themselves at older ages by change their food, environment, even the pace of living. Also that new approach of prediction pre-cancer will help if there is any treatment can give for that person to therapy the mutated tumor protein P53.

—————————— ◆ ——————————

## 1 INTRODUCTION

TP53 gene provides instructions for making a protein called tumor P53 protein. This protein as a tumor suppressor, which means it regulates cell division by keeping cells grow and divide very rapidly or in an uncontrolled manner. P53 tumor protein is located in the nucleus of cells throughout the body, where it connects directly to DNA.

TP53 gene (Cytogenetic) Location: 17p13.1, Molecular Location on chromosome 17: base pairs 7,571,719 to 7,590,867. Fig. 1 shows the TP53 gene is located on the short (p) arm of chromosome 17 at position 13.1. More precisely, the TP53 gene is located from base pair 7,571,719 to base pair 7,590,867 on chromosome 17 [1].

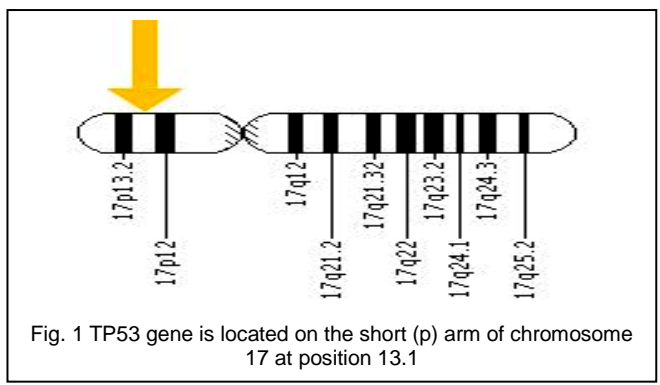

Fig. 1 TP53 gene is located on the short (p) arm of chromosome 17 at position 13.1

————————————————

• Ayad Ghany Ismaeel is currently pursuing Professor Assistant, PhD computer science in department of Information System Engineering, Erbil Technical College- Hawler Polytechnic University (previous FTE- Erbil), Iraq. PH-009647703580299. E-mail: dr_a_gh_i@yahoo.com Alternative dr.ayad.ghany.ismaeel@gmail.com

1979 the P53 gene, later found to be the most frequently mutated gene in human cancer, is discovered. Since the discovery of a gene human due to tumor cancer in 1981, called oncogene, primary cause of cancer is the occurrence of mutations in specific genes (the important one TP53 gene which caused of cancers), as a result of exposure to toxins or radiation or by processes corrective wrong of DNA or the result of errors that occur when transcription of DNA before cell division, addtional the advance in old may play role at exchange genatic cells means the DNA and the RNA reached to turom protien P53 and in the final causes the cancers [2]. There are factors due to injury with cancer and distortion material genetic such as distortion physical factors such as gamma-ray ionizing liberated from the nuclei of atoms of heavy elements or high-energy particles such as alpha or beta particles and high temperatures. In addition to chemical factors such as the concentrations of chemical compounds that interfere with the body either through food or drink, breathe, or even some mechanical factors such as an impressive blow against the body causes of the machine lead to inflammation and the decline of blood or other and these are the environmental impacts. And cancer is a tumor and cirrhosis caused by cancer cells.

Well known that with the aging increases the risk of cancer and the reason is mainly the result of accumulated hurt the DNA with the passage of time. A new study suggests another reason for the increased incidence of cancer in the elderly, and is diminishing the effect of the gene responsible for the fight against cancer TP53. Where named P53 gene is one of the most important genes which operates inhibitor to cancer, which makes the cell response to repair or suicide cell, otherwise will cause cancers, and it found more than 50% of cancer patients have mutations in the gene P53 [3].

The changes in the TP53 gene related with cancers (skin, liver, lung, bladder, breast, head and neck, esophagus, stomach and colorectal cancers, and hematological malignancies), as follow [1, 4]:

A. Breast cancer - increased risk from variations of the TP53 gene: Changes in the TP53 gene greatly increase the risk of

developing breast cancer as part of a rare inherited cancer syndrome called Li-Fraumeni syndrome. These inherited mutations are thought to account for less than 1 percent of all breast cancer cases. Somatic mutations in the TP53 gene are much more common, occurring in approximately 20 percent to 40 percent of all breast cancer cases. Many of these mutations change a single protein building block (amino acid) in tumor protein p53. These mutations lead to the production of a nonfunctional version of this protein.

B. Bladder cancer - associated with the TP53 gene: Some gene mutations are acquired during a person's lifetime and are present only in certain cells. These changes, which are called somatic mutations, are not inherited. Somatic TP53 mutations in bladder cells have been found in some cases of bladder cancer. Most of these mutations change a single protein building block (amino acid) in tumor protein p53.

C. Li-Fraumeni syndrome - associated with the TP53 gene: Although somatic mutations in the TP53 gene are found in many types of cancer, Li-Fraumeni syndrome appears to be the only inherited cancer syndrome associated with mutations in this gene. More than 60 different mutations in the TP53 gene have been identified in individuals with Li-Fraumeni syndrome. These mutations are typically inherited from a parent and are present in all of the body's cells. Many of the mutations associated with Li-Fraumeni syndrome change a single protein building block (amino acid) in the part of tumor protein p53 that binds to DNA.

D. Other cancers - associated with the TP53 gene: Somatic mutations in the TP53 gene are the most common genetic changes found in human cancer, occurring in about half of all cancers. For example, TP53 mutations have been identified in several types of brain tumor, colorectal cancer, a type of bone cancer called osteosarcoma, a cancer of muscle tissue called rhabdomyocarcinoma, and a cancer called adrenocortical carcinoma that affects the outer layer of the adrenal glands (small hormone-producing glands on top of each kidney).

Most TP53 mutations change single protein building blocks (amino acids) in tumor protein p53, which leads to the production of an altered version of the protein that cannot bind effectively to DNA. This defective protein can build up in the nucleus of cells and prevent them from undergoing apoptosis in response to DNA damage. The damaged cells continue to grow and divide in an unregulated way, which can lead to cancerous tumors.

The big problem happens when the change at TP53 will product type of P53 protein can't work normally, i.e. it can't regulates cell division by keeping cells grow and divide very rapidly or in an uncontrolled manner, moreover defective protein can build up in the nucleus of cells and prevent them from undergoing apoptosis in response to DNA damage. As results foreword their high risk of developing cancers in breasts, bladder, brain, osteosarcoma etc. Whenever the person can diagnosis and predict the change at P53, that will make the person can take steps to reduce their risk of developing or lead to cancerous tumors, while for persons which holder to the mutated gene causes the cancer disease, in this case the person can monitor changes in breasts, bladder, brain, etc carefully to find cancer at an earlier, more treatable stage [4]. So needed an approach to diagnosis and predict to find the product P53 protien is normal or abnormal, means pre-cancer technique via detection mutations change at single protein building blocks (amino acids) in tumor protein P53.

## 2 RELATED WORK

Jonas Carlsson, Thierry Soussi, and Bengt Persson [2009], a method has been developed to predict the effects of mutations in the p53 cancer suppressor gene. The new method uses novel parameters combined with previously established parameters. The most important parameter is the stability measure of the mutated structure calculated using molecular modelling. For each mutant, a severity score is reported, which can be used for classification into deleterious and nondeleterious. Both structural features and sequence properties are taken into account. The method has a prediction accuracy of 77% on all mutants and 88% on breast cancer mutations affecting WAF1 promoter binding. This study done investigated of correlations between human p53 mutants found in cancer patients and the corresponding activity of promoter binding. The aim was to obtain a better understanding of molecular mechanisms to explain why certain mutations cause more severe effects than others and to be able to predict the severity of new, hitherto uncharacterized mutants [3].

Ichikawa A., Hotta T., Takagi N., and others [2011], the alteration of p53 tumor suppressor gene was studied in 48 patients with 8-cell lymphoma. A sequential combined technique of polymerase chain reaction-mediated single strand conformational polymorphism (PCR-SSCP) or reverse transcription (RT)-PCR-SSCP and direct sequencing were used as a simple and sensitive approach to analyze nucleotide changes. By these methods, we identified 8 missense point mutations and 2 codon deletions in 9 of the 48 patients. These mutations were located in or close to the evolutionarily [5].

The drawbacks of these methods and techniques focus in diagnostic or prediction base on features of mutations in disease's genes not in genome sequence and for patients have cancer and not pre-cancer. The motivation overcome the drawbacks of the previous techniques to reach a new approach for prediction pre-cancer via detection mutated in tumor protein P53, by considering and introducing multi other features show the alternations, changes in the environment as well as TP53 gene, comparing sequences to gain information about the structure/function of a query sequence. Also proposing optimal and more accurate approach for classification and dealing with specific TP53 database and taking into consideration depending on the idea of "two sequences may have big differences in DNA sequence but have similar protein" [6, 7]. So will not stop at TP53 gene sequence but will considered the mutations at tumor protein P53 protein of TP53) which is play important role as a tumor suppressor.

## 3 PROPOSED OF NEW APPROACH FOR PREDICTION PRE-CANCER VIA DETECTION MUTATED IN P53

The survival rate is high with early diagnosis, 97% of women survive for 5 years or more years, Predicting malignancies plays essential roles not only in revealing human genome, but also in discovering effective prevention and treatment of cancers, curing from this malicious disease is mainly based upon early and accurate diagnosis computer aided diagnosis can be a good helper in this area [6].

The new approach of prediction pre-cancer via detecting mutated in tumor protein P53 will base on bioinformatics tools using normal TP53 gen and its protein [6], after satisfying that normal homo TP53 gene is represented to the environment by checkup GC% content [7]. The important feature needed with this proposed approach the TP53 database used to classify and diagnosis the malignant mutations at tumor protein P53 because this protein at level of certain codon will see there is multiple malignant mutations caused to multiple types of cancers can not predicting and detection them without TP53 database, and the better select which related to its' environment. The proposed algorithm of new approach shown as follow:

---

**Algorithm of new approach for prediction pre-cancer via detecting mutated in tumor protein P53**

**Input:** DNA sequence (normal and person's gene sequence).
**Output**: Diagnose and predict there are malignant mutations of person's tumor protein P53 or not (No risk).
**BEGIN**
  **Step 1:** Get the TP53 person's gene sequences
  **Step 2:** Put that TP53 person's gene sequences in FASTA format;
  **Step 3**: Using BioEdit package to search about the normal homology TP53 gene using BLAST NCBI
  **Step 4:** Calculate the GC% and AT% of normal homology TP53 gene using BioEdit package
  **Step 5:** if GC% equal or greater than 38% then
         Depend it as normal PT53 gene for test person's TP53 gene
     else
         Find another normal homology TP53 gene seguence form another database like EBI, Ensemble, etc and return to step 4
  **Step 6:** CREATE fasta file contains the depended normal homology TP53 gene and person's TP53 gene;
  **Step 7:**  SET fasta file to clustalW for alignment; display alignment result
  **Step 8:** if there is mutations
          Translate the normal TP53 gene and the person TP53 gene to proteins using BioEdit package.
        if there is mutations   // Classification there is mutations
             Sereach T53 database for matching (pre-cancer dedection)
             if there is matching
                 Retrieve the biologic references of the cancer (types, names, positions, etc)
             else
                 unknown cancer
        else
             No risk
     else
         No rsik
**END**

---

## 4 EXPERIMENTAL RESULTS
### 4.1 Bioinformatics Techniques

The implement and apply of new approach for prediction pre-cancer using BioEdit ver. 7.2.0 as shown in the following steps:
1. Obtain the person's TP53 gene sequence from oncogene labs will use it in FASTA format to reach the normal homology of TP53 gene using BioEdit package.
2. Reach to Normal homology TP53 gene using BioEdit by select World Wide Web→ BLAST at NCBI→ nucleotide blast→ Paste or upload from file the person's TP53 gene sequence which is formatted in FASTA as shown in Fig. 2.

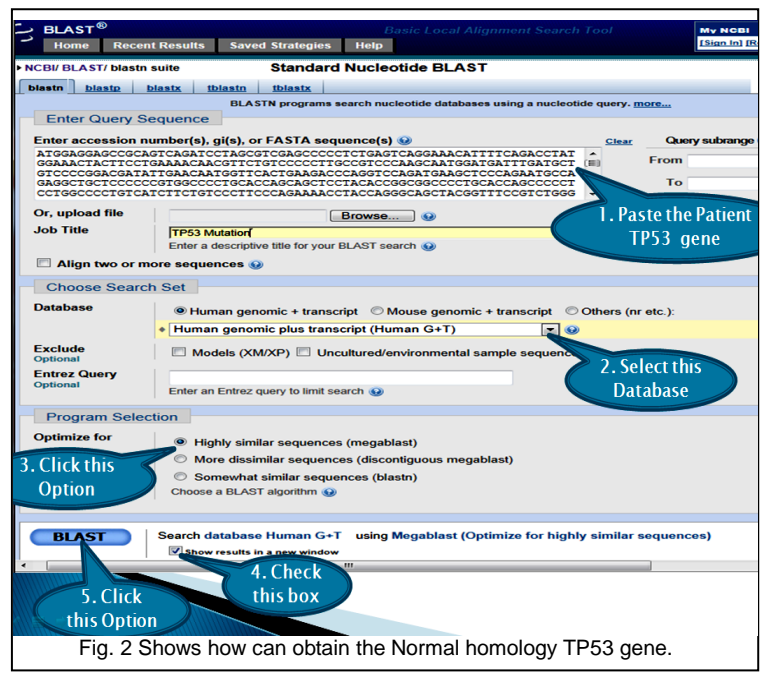

Fig. 2 Shows how can obtain the Normal homology TP53 gene.

3. This step gives two options to get the normal TP53 gene sequence as shown in Fig. 3. When click one of them will reach to normal homology TP53 gene as shown Fig. 4.

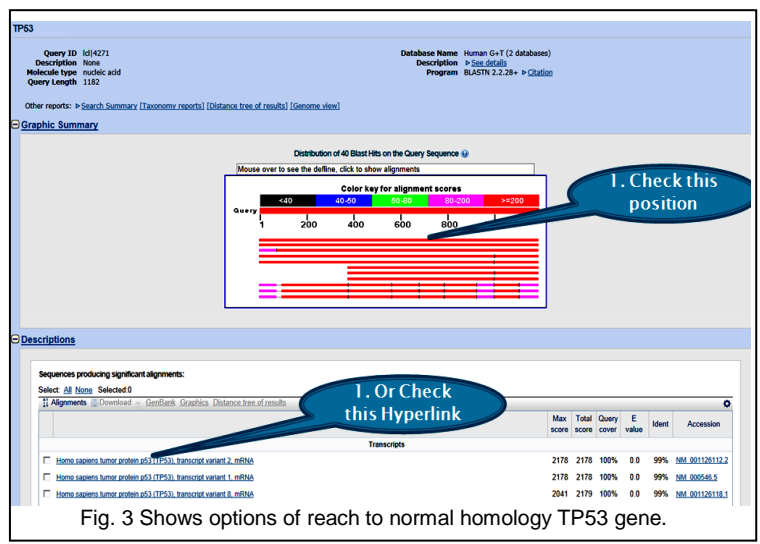

Fig. 3 Shows options of reach to normal homology TP53 gene.

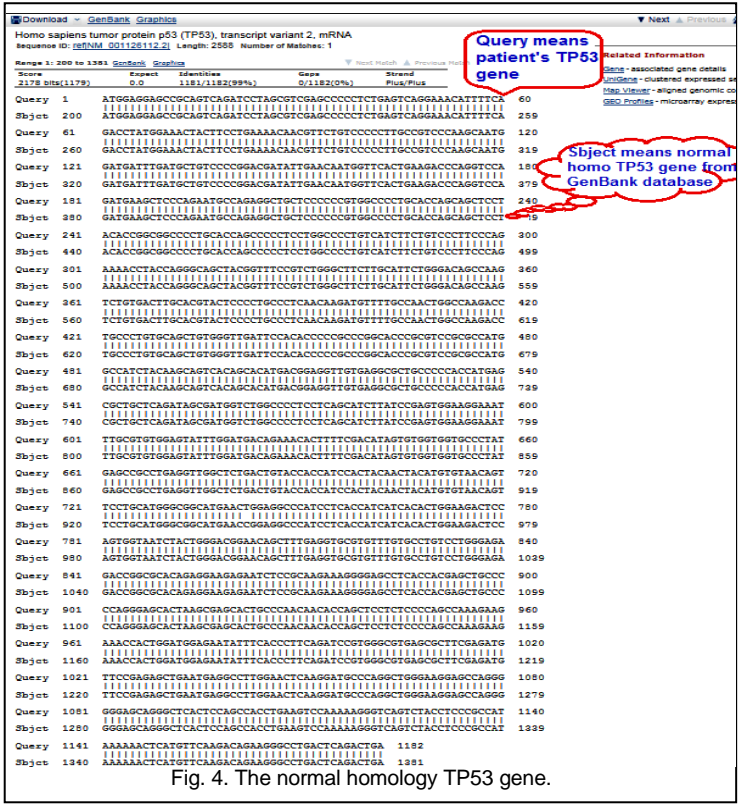

Fig. 4. The normal homology TP53 gene.

4. Open the BioEdit package with the normal gene of TP53 (Fasta file) which obtained in step (3) to calculate the GC% content by select sequence option→ Nucleic Acid→ Nucleotide Composition, that will give the GC% whether equal or greater than 38%, Fig. 5 shows it satisfy 54.85%, so will go to the next step, otherwise will return to step 2 to find another database to reach from it to the normal TP53 gene satisfying GC% >= 38%, and then continue with the same step 3, step 4 etc.

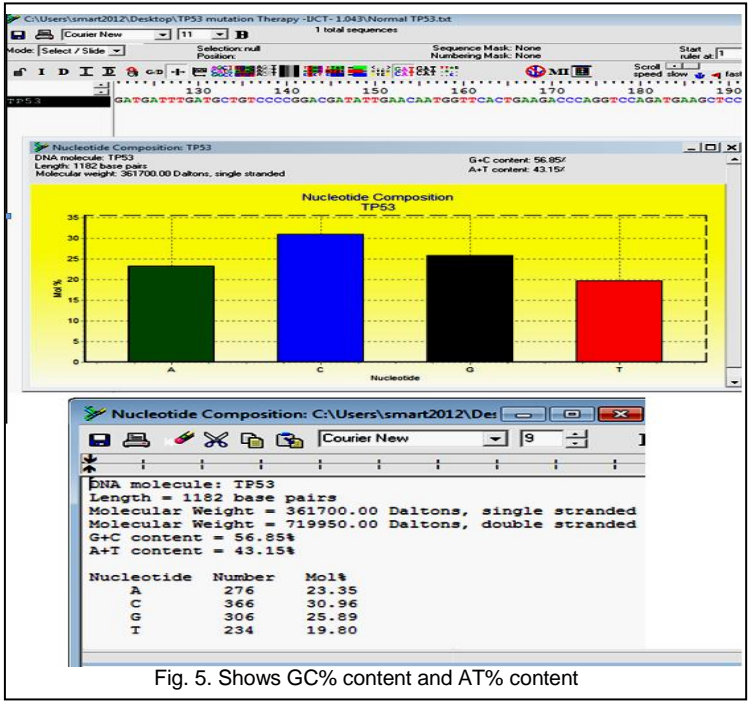

Fig. 5. Shows GC% content and AT% content

5. Create fasta file contains the selected normal homology TP53 gene and the person's TP53 gene.
6. Using that fasta file which obtained in step (5) to clustalW to display alignment result, i.e. diagnosis there is malignant mutations by comparing the normal TP53 gene sequence with one (or more than one) person's TP53 gene sequences at the same time. That is done by BioEdit will select Accessory Application→ ClustalW Multiple Alignment→ Run ClustalW, then obtained the result whether there is malignant mutation or not, and in this example will see mutation at person's TP53 gene comparing with normal TP53 gene as shown in Fig. 6.

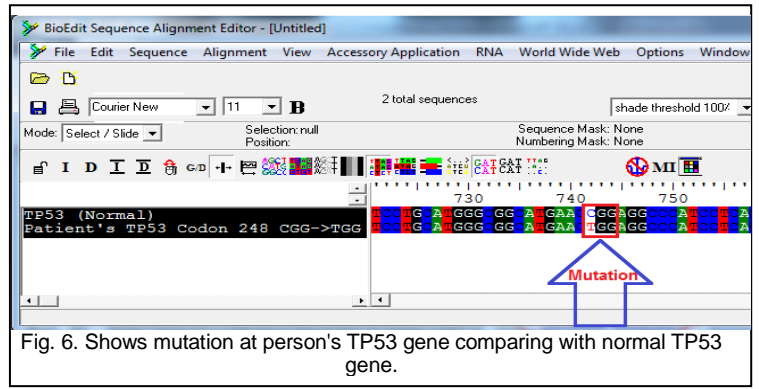

Fig. 6. Shows mutation at person's TP53 gene comparing with normal TP53 gene.

7. That not enough so needed to transform to tumor protein P53 of normal homology TP53 gene and P53 protein to person's TP53 gene then using the same tool ClustalW at BioEdit package to diagnosis there is malignant mutations as done in step (6) or not (No risk), Fig. 7 shows there is malignant mutations at codon 248 (CGG→ TGG), i.e will find the codon 248 converted from R (at Normal P53) to W (at Person's P53 gene)**.**

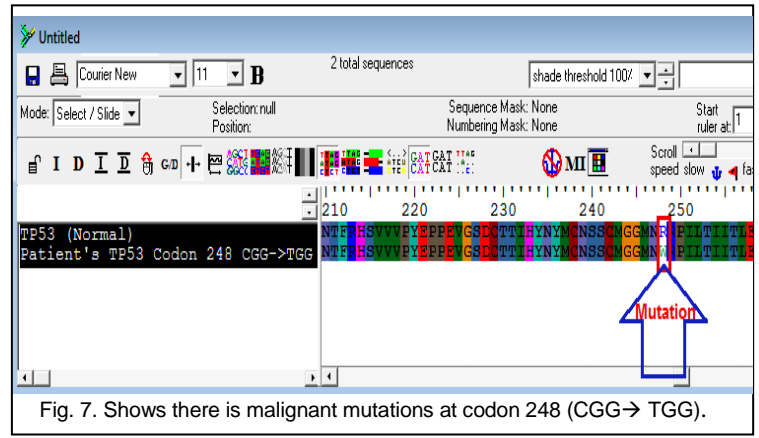

Fig. 7. Shows there is malignant mutations at codon 248 (CGG→ TGG).

8. This step used to classify that malignant mutations which are discovered at step (7) to predict pre-cancer for persons by determine the cancer(s) may be related to these mutations which expected there is multiple types related with each mutated codon in tumor pro-tein P53, e.g. the mutation in codon 248 will need database of TP53 (P53) gene and is better related with environment (locally database represent the persons, means if the person from Asia better to be this database of TP53 from Asia, more specific in Middle east, or national database, etc.). For that resons to implement and applying the proposed approach will select the database (UMD_Cell_line_2010) of TP53 website as modern and comprehensive datatbase (URL: http://p53.free.fr/Database/p53_MUT_MAT.html) [8] , can use any other locally database related to environment because it become more representative in the case of pre-diagnosis and prediction of cancer. Fig. 8, explain how can get full related information like the insert, update, delet, etc of mutation which is discovered at person's P53 gene. Addition can obtain very specific oncogene laboratories information related to mutated codon (248) by implement more than one filter for the certain fields like Codon, Mutated codon, Cancer, etc. Fig. 9 shows example using this TP53 database, how can obtain more specific results by doing the filters using more attributes to codon=248, Mutated codon= TGG, and Cancer= Colorectal carcinoma respectively, and increasing the attributes make reached to more specific results.

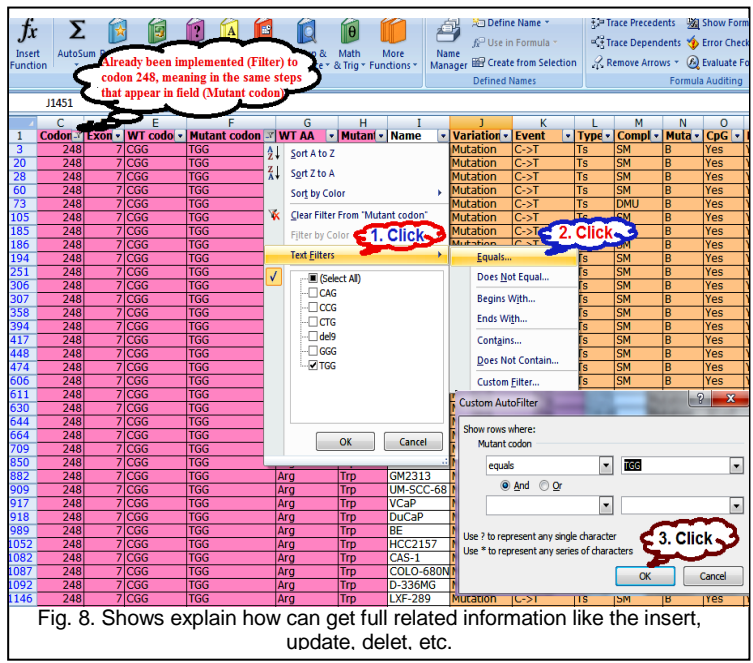

Fig. 8. Shows explain how can get full related information like the insert, update, delet, etc.

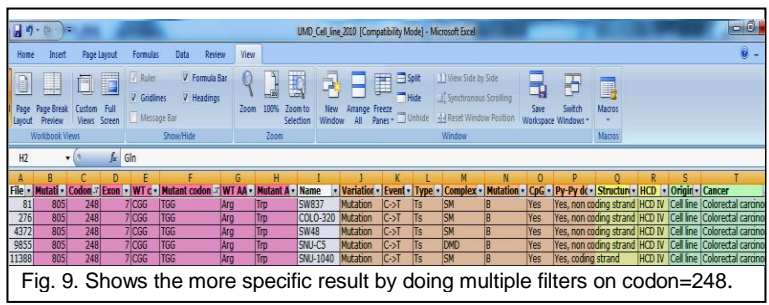

Fig. 9. Shows the more specific result by doing multiple filters on codon=248.

The results of pre-cancer prediction which obtained from Applying the proposed approach shows multiple cancers may be caused by certain codon, Therefore, even among patients with cancers of the same body organ or tissue, the genetic profile of each individual's tumor can differ greatly [2], e.g. codon=248 gives various cancers as shown in Fig. 8, and certain cancer in different structures as shown in Fig. 9. Sometime may reach to nothing, but in the cases of there is malignant mutation but unknown cancer or the effect which is caused. In the case of diagnosis multiple cancers must be using multiple features include the mutated codon which caused the cancer to extract the specific results that will help to detect and diagnosis pre-cancer(s) (prediction) which may infect that person in future.

### 4.2 Discussion the Results

Table 1 show comparing the results of proposed approach for prediction pre-cancer via detecting mutated P53 with other techniques or methods.

TABLE 1
REVEALS COMPARISON OF PROPOSED APPROACH WITH OTHER TECHNIQUES OR METHODS

| Proposed Approach for Prediction Pre-cancer via Detacting Mutated P53 | A Ichikawa, T Hotta, N Takagi, and others [5] | Jonas Carlsson, Thierry Soussi, and Bengt Persson [3] |
|---|---|---|
| Proposed approach for Prediction pre-cancer | approach diagnose and monitoring after infecting cancer | approach diagnose and monitoring after infecting cancer |
| Diagnose and Detecting malignant mutations based on datasets sequences of Normal TP53 gene and person's TP53 gene with their tumor proteins P53. | It works direct with Person's TP53 gene and its tumor protein P53 only | It works direct with Person's TP53 gene and its tumor protein P53 only |
| Cost-effective for analysis where the cost about $3000 (cost of obtain P53 sequence from oncogene labs[6] | Does not have this Cost-effective | Does not have this Cost-effective |
| Considered the GC% and AT% content to select the Normal TP53 gene. | There isn't | There isn't |
| Classifying and diagnoses of malignant mutations at P53 needed TP53 database or TP53 website because there is a large number of malignant mutations and types of cancers associated with P53. | There isn't | There isn't |
| This proposed approach friendly to use by researcher, molecular biologic, Biomedical, etc | Limited | Limited |

## 5 CONCLUSIONS

The proposed approach of prediction pre-cancer detecting via mutations disease at tumor protein P53 shown the following conclusions:

A. The proposed approach offers friendly diagnosis and detecting malignant mutations and of pre-cancers as shown in Table 1, i.e. can use by researcher or any person who needed to test malignant mutations at TP53 gene and its' protein (P53) which caused more than 50% human cancers.
B. This new approach is more effective when used query of TP53 database with multiple attributes/features to extract specific result as shown in subsection 4.1; step 8, i.e. when the user increase the number of filters this approach will give more specific result, and vice versa.
C. The proposed approach shows the detecting pre-cancer via predicting mutated P53 gene reducing of infect in cancers by avoiding exposure to toxins, radiation or when monitoring themselves at older ages and change their food, environment, even their living, addition if there is any treatment can give for that person to therapy the mutated tumor protein P53.
D. The proposed approach of prediction pre-cancer must be based on collection datasets (normal TP53 gene and its tumor protein P53), as well as the TP53 database if exist which used for diagnosis and classify the type of malignant mutations and the cancer may caused in tumor protein P53 related to environment that means the prediction is more expressive for the region and will be better as this as have seen from the above impact on the results and how it could affect the environment.

## ACKNOWLEDGMENT

Thank to my wife (DR. NEMA SILAH ABDAL KAREEM) that helped me in the maturation of the idea of research and advices at the medical side, which has to do with their competence.